# Responding to Challenge Call of Machine Learning Model Development in Diagnosing Respiratory Disease Sounds

*Corresponding Author:*

**Negin MELEK (Negin MANSHOURI)**

Avrasya University

Faculty of Engineering

Department of Electrical and Electronics Engineering

61080 Trabzon/TURKEY

Phone (mobile): (+90) 5535791640

E-mail: negin.melek@avrasya.edu.tr

# Responding to Challenge Call of Machine Learning Model Development in Diagnosing Respiratory Disease Sounds

*This paper is dedicated to the memory of the late Jalal and Behzad Manshouri, who passed away from COVID-19 in September 2021.*

*Abstract*

The normal and abnormal sounds arising from the respiratory system shed great light on the medical science world by revealing the quality, diseases and changes in the lungs of people. In medicine, this invasive and easy old method, which is realized by the stethoscope, facilitates the diagnosis of diseases by specialists. This manual method can sometimes lead to wrong decisions in terms of sound detection due to different audibility. In lung diseases with a high mortality rate, detailed sound analysis is very important for obtaining accurate detection. As technology advances, the development of automated approaches based on machine learning is of great interest as they provide modern and highly accurate analysis. In today's most popular topic, i.e., the COVID-19 disaster, the conflict of early detection of respiratory disease and machine learning for sound signal processing is of great interest. In this study, a machine learning model was developed for automatically detecting respiratory system sounds such as sneezing and coughing in disease diagnosis. The automatic model and approach development of breath sounds, which carry valuable information, results in early diagnosis and treatment. A successful machine learning model was developed in this study, which was a strong response to the challenge called the "Pfizer digital medicine challenge" on the "OSFHOME" open access platform. "Environmental sound classification" called ESC-50 and AudioSet sound files were used to prepare the dataset. In this dataset, which consisted of three parts, features that effectively showed coughing and sneezing sound analysis were extracted from training, testing and validating samples. Based on the Mel frequency cepstral coefficients (MFCC) feature extraction method, mathematical and statistical features were prepared. Three different classification techniques were considered to perform successful respiratory sound classification in the dataset containing more than 3800 different sounds. Support vector machine (SVM) with radial basis function (RBF) kernels, ensemble aggregation and decision tree classification methods were used as classification techniques. In an attempt to classify coughing and sneezing sounds from other sounds, SVM with RBF kernels was achieved with 83% success.

## 1. Introduction

The respiratory system, which is the most important feature of being alive, helps us breathe by including the airways, blood vessels and lungs. Owing to the cooperation of the muscles that strengthens the lungs and the respiratory system, gas exchange takes place in the body [1]. It is surprising that the respiratory system has many important functions besides helping with breathing, which include the ability to speak and smell, balance air temperature and body temperature as well as remove harmful substances and waste gases from the body. By considering the factors affecting this vital organ, great success can be achieved in the early diagnosis of respiratory tract infections and diseases. The sounds emerging from the lungs have a great role in the early diagnosis of respiratory system diseases [2], [3], [4], [5]. Intensity, frequency and quality, which are the characteristics of these sounds, are the considerations in distinguishing similar sounds from each other [6].

The presence of cough symptoms in different types of respiratory diseases and the point that this symptom is a useful tool in the diagnosis of the disease have been the favorite subject of many studies [7]. The reflex that defends the lungs against any irritant in the respiratory system is coughing. Most of the time, self-healing cough can be a sign of an important disease in some cases. Cough, which is a symptom of many diseases such as COVID-19, bronchitis, lung cancer and asthma, has different characteristics in every disease in terms of sound and provides great convenience to doctors as to diagnosis [8], [9]. In addition to the field of medical science, cough automatic sound analysis and early diagnosis of respiratory diseases have been the main target of many research. Thus, automatic analysis and classification work of cough or sneeze, which is a symptom of deadly diseases such as COVID-19 in respiratory diseases, has become important today. Artificial intelligence, which provides great convenience and prosperity in people's lives, has become a technology that gives promising results in many studies [10], [11].

Studies in the field of early diagnosis of the cough-based disease can generally be categorized into two groups. While the first group is the cough sound classification in a dataset containing different sound types, the second group is to classify the cough types.

The presence of sputum in the lung in the dry and wet cough type classification was extensively detected in [12]. Sputum detection should be considered as the first sign of many diseases such as pneumonia, cancer and infection. In the clinical settings, sputum detection examinations are performed individually. This study, which increases the accuracy of these tests and facilitates this detection, proposes a different way according to the characteristics of cough sounds. Dry and wet cough sounds from 131 participants were analyzed as a multi-layer labeling platform. As a result, 88% sensitivity and 86% specificity were achieved in dry and wet cough classification.

In 2013, a valuable and pioneering study developed an automatic and early detection model for pediatric pneumonia based on the analysis of respiratory system sounds [13]. This disease, which has a high mortality rate due to the lack of laboratories and the small number of health teams in poor areas, has led to promising results owing to not requiring any physical contact. Successful decisions were obtained by extracting effective features from cough sounds and applying logistic regression classifier. Based on these effective features, the targeted disease could be differentiated from other diseases with 94% and 75% sensitivity and specificity, respectively.

A major advantage of automatic cough sound type classification studies is to minimize the error rate in the detection of dry or wet cough type, which is based only on the subjective judgments of doctors [14]. To largely solve this problem, a model for sound type analysis has been developed. After the signal processing steps were passed, the results obtained by automatic classification were compared with the decisions determined by the two experts. As the result of this comparison, it was concluded that the proposed model is a useful tool for cough-based remote disease monitoring and diagnosis.

In the literature review, studies based on different cough classifications are found. The main goal of these studies is to distinguish the cough sound from among several sounds. Detailed analysis of cough frequency and severity in the patients suffering from cough as the result of chronic diseases provides valuable information. Based on this issue, automatic cough sound detection is made from the recordings taken via mobile using the hidden Markov model [15]. Based on the results of the proposed model, the feasibility of the hidden Markov method in detecting cough in mobile patients is demonstrated.

In another study conducted in 2019, information about preprocessing in cough sound detection, especially in noisy environments, was presented. This study showed that a preprocessing step was necessary to suppress the noise in cough sounds in order to minimize the margin of error in the diagnosis of respiratory disorders [16].

A model proposal that detected abnormal situations by examining cough sound information was made in [17]. The main purpose of this real-time model was to provide remote monitoring of older and lonely individuals and early intervention in critical situations. Two models were used for cough sound classification in the dataset, which included different environmental sounds as well as cough. One of them was neural network and the other was hidden Markov model. This model was shown to provide high performance at a low signal-to-noise ratio. It is noteworthy that by presenting a simple prototype of the proposed model, it provided great convenience to the user candidate due to its use of the wireless microphone.

For the detection of different patterns in the flow of cough sound, a system using an acoustic detector was presented in [18]. The study was focused on the classification process to distinguish the impulse patterns contained in the cough sound from other impulsive sounds. This system, which is strong against noise and reverberation, showed 90% and 99% sensitivity and specificity, respectively, due to having long-short-term-memory architecture used in the field of deep neural networks.

The difficulties and advantages of studies with artificial intelligence methods in the field of cough sound detection and early disease diagnosis were presented in [7] in the form of a review study. In this comprehensive review based on the cough sound classification, the use of different methods was compared. The most commonly used method in cough-based disease diagnosis studies was chosen as logistic regression and support vector machine. On the other hand, artificial intelligence algorithms compared with random forest algorithm and deep learning architecture were widely preferred.

In this study, an automatic early detection system based on coughing and sneezing sound classification was tried to be modeled. In our work, we focused on the design of a new machine learning model, paying close attention to the invitation of the open Science Center "OSFHOME" platform [19]. The three-part dataset created by this platform for the "Pfizer digital medicine challenge" consists of training, validating and testing sets. As the result of the efficient feature extraction method from this dataset, the classification stage was entered to complete the designed machine learning model. The validation dataset was analyzed to obtain the parameters of the used classifier algorithms.

In this dataset, the class covering coughing and sneezing sounds could be distinguished from the class comprising other sounds. The designed machine learning model could be useful in the early diagnosis of the current widespread COVID-19 disaster, as well as other diseases caused by viruses that spread through coughing and sneezing in crowded places.

## 2. Methods and test protocol
### 2.1. Dataset

The dataset of the proposed article was created from ESC-50[20] and AudioSet [21] audio files. The general introduction of these datasets can be presented briefly and concisely in the following sentences: The ESC-50 dataset is known as a collection of approximately 2000 types of labeled sounds suitable for sound classification studies. This dataset includes five major categories as the sounds of animals, natural sounds/water sounds, human sounds excluding speech, domestic sounds and city noise sounds. ESC-50, one of the popular datasets since 2016, has many studies in terms of audio signal analysis and classification [22], [23], [24].

In terms of data usability, the fast access, up-to-date and reliability of OSF creates a solid basis for researchers to conduct projects. The open access dataset in the presented article can be accessed from the link below:

https://osf.io/tmkud/

By making use of artificial intelligence algorithms based on machine learning in the age of technology, this dataset has the ability to play an important role in the early detection of respiratory tract infections, as well as in the analysis of different respiratory sound-induced diseases.

The AudioSet dataset, consisting of YouTube videos, is composed of 10-second human-labeled audio clips. A detailed study has been made on the AudioSet dataset in order to open a door to acoustic sound event detection [25].

In the proposed coughing and sneezing sound classification analysis, the graph representing the training, validating and testing sample number in the three-part dataset consisting of ESC-50 and AudioSet audio files is presented in Fig. 1. As can be seen, there were 3718, 1221 and 1654 samples, respectively, for training, validating and testing.

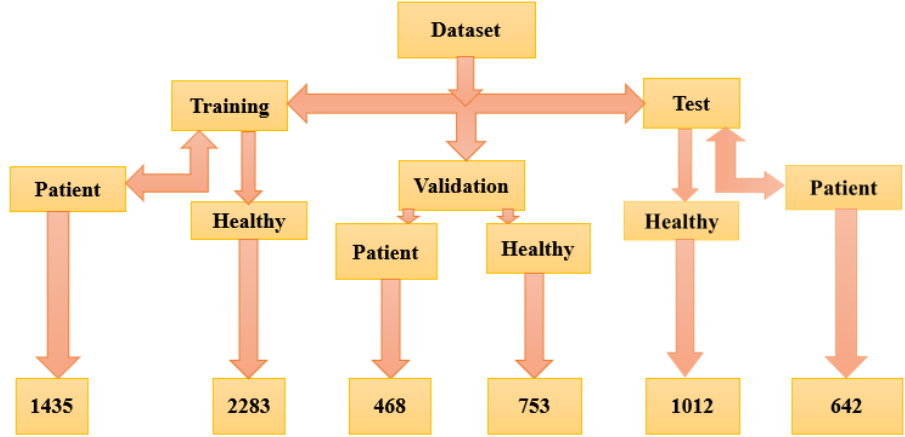

Fig. 1. Sample distribution of the three-part dataset.

In the study, the dataset was divided into two groups as patients and healthy individuals. The class containing the patient labels included respiratory diseases such as sneezing and coughing, while the healthy class included common human-induced sounds such as laughing and singing. Sound recordings with the length of 4.98 seconds were sampled at the frequency of 44100 Hz. All the data analyses in the study were performed in Matlab 2019b application.

*2.2. Data processing*

*2.3. MFCC feature extraction technique [26]*

In the analysis of audio signals, MFCC feature extraction is used as an efficient and common method [27], [28], [29]. Different techniques such as linear prediction coefficients (LPC) and perceptual linear prediction (PLP) coefficients are also used as feature extraction methods. MFCC has proven to be more successful than other methods in speech recognition systems [29].

This technique generally consists of five basic parts. These are respectively pre-emphasis, signal framing and windowing, applying the discrete Fourier transform, calculating the logarithm of the magnitude and multiplying the frequencies to the Mel scale (called the Mel filter bank step) and, finally, computing the inverse discrete cosine transform [30], [31]. The first stage, pre-emphasis, is actually a process that compensates for the rapidly distorting spectrum of the audio signal. Framing is the other step used to divide the audio signal into smaller compartments [32]. The most important task of the windowing process is to prevent discontinuity of the obtained audio signal [33]. In this study, the windowing process was performed by selecting the Hamming window. The working principle of this window is shown in Equation (1):

$$W[n] = 0.54 - 0.46 \cos\left[\frac{2\pi n}{N-1}\right] \qquad (1)$$

$W[n]$ and $N$ are the $n^{th}$ coefficient of the Hamming window and the number of samples per frame, respectively [33]. Following the windowing process, the Fourier transform is used to move the audio signal from the time domain to the frequency domain. The other stage, the Mel filter bank, consists of overlapping filters. Human ear structure and MFCC technique are compatible and similar in terms of frequency band change [34]. In this experiment carried out by the researchers in 1940, they showed that the frequencies of the auditory system were linear up to 1kHz and that they underwent logarithmic variation at higher values [35]. In this context, there was a relationship between the actual frequency and the perceived frequency known as Mel, as presented in Equation (2):

$$f_{Mel} = 2595 \log_{10}\left(1 + \frac{f}{700}\right) \qquad (2)$$

In this equation, $f_{Mel}$ is the output of the filter bank and $f$ is its input. 2595 and 700 in the equation are constant numbers [36].

*2.4. Sequential forward selection (SFS) as feature selection method*

Feature selection (FS) is a useful process used to reduce the number of obtained features and select more efficient ones. In the problem-solving phase, some unnecessary features cause confusion as well as a cluttered feature space and decrease the classification performance [37]. FS is of great benefit in machine learning studies as simplicity of operation, accuracy and speed [38].

There are many different methods for reducing the feature set size and selecting dominant features. Among these methods, it is known that the sequential forward selection (SFS) technique is successful in terms of speed and easy understanding [37], [39], [40].

This algorithm is based on sequential feature selection. Working as a bottom-up search tool, this algorithm gradually adds features that seem suitable as the result of the computational functions from an empty set [41], [42]. An important advantage of this algorithm is that the newly chosen feature is selected from the remaining feature set, so that the newly designed and expanded set will have a minimum error in terms of the classification process compared to other additions [38].

*2.5. Classification algorithms*

*2.5.1. Nonlinear SVM and RBF kernel functions*

The SVM classification method [43] is a frequently preferred algorithm in regression analysis and classification-based research. In machine learning and signal processing, this supervised model is known as a robust method in the prediction field due to its strong statistical basis and ability to minimize the probable risk ratio [44]. The easiest and most understandable definition of the SVM method can be expressed as follows: it reveals a decision hyperplane by considering the optimum support vectors and then performs the most appropriate data classification in the dataset. Thus, it is an ideal choice for two-class problems [31], [45]. This algorithm, which has its roots in the 1900s [46], was first shown to be capable of linearly separating samples in a dataset with a linear format. Thus, in this case, it is responsible for choosing the hyperplane that maximizes the margin between the two classes with linearly distributed samples [47].

A nonlinear SVM classifier can be obtained as the result of a nonlinear operator application called 'kernel trick'. Owing to this trick, data analysis can be moved to the multidimensional feature space [48]. One of the important factors affecting the classification result is the selection of kernels and effective parameters related to these kernels [49]. In the study, the most common RBF kernel in the SVM classification method was used. The parameters that directly affect the performance of this nonlinear SVM type are $C$ and $\gamma$. Parameter $C$ is actually defined as a regulation criterion in SVM. For the sake of maximizing the margin of the decision function, this parameter sometimes prevents the correct classification of the samples. In the hyperplane representation, a limited and few selected points are the result of a small $C$ value in determining the critical boundary between the classes. This wrong choice may result in obtaining incomplete information [50]. On the other hand, with a large $C$ value, wrong decisions can be made again by obtaining a large decision limit due to the selection of more sample points. As a result, choosing the optimum value of $C$ can minimize the error in the training phase [49]. Another important parameter in the designed training model is $\gamma$. This parameter actually shows the desired degree of curvature in the decision boundary. For two classes, the optimal selection of the gamma value, which shows the distance between the decision boundary and the nearest support vectors, is important.

A high $\gamma$ value means choosing the ones closest to the decision boundary in terms of samples and a low $\gamma$ means choosing the farthest points on the decision boundary. RBF, which is one of the kernel varieties and selected in the study, is presented in Equation (3). In the presented formula, σ shows the standard deviation (SD) of the samples and is chosen as 1 in this study.

$$K_{RBF}(x,y) = \exp\left(-\frac{\|x-y\|^2}{2\sigma^2}\right) \tag{3}$$

*2.5.2. Decision tree*

Decision tree, which is one of the important classifiers in artificial intelligence based on machine learning, has high speed, powerful learning model and an easy structure [51], [52]. This algorithm, which falls under the category of supervised classifier, is used to solve regression and classification problems. The purpose is to create a model to predict the class of an unknown sample based on the decision rules learned from the training data. This method, which subdivides the dataset of analysis, consists of the root nodes and internal nodes. Each node is formed of a single main part and two or more sub-parts called descendants [53]. Based on Magerman's view, the decision tree is applied to the model to support the decision process at the classification stage. Thus, it creates possibilities in each choice for the different states of the decision [54].

The first node in the decision tree is the root node. As the result of the root state evaluation, the observation process that results in "Yes" or "No" is classified. Internal nodes representing attributes are located below the root node. As the number of nodes increases, the complexity of the designed model will increase. In this tree-like flowchart, the part that presents the result is the leaves, which are the lowest nodes [55]. This method, which is compatible with different types of variables, has great advantages in the field of machine learning because it is easy to understand and can classify with optimum calculation. The "fitctree" command was selected to set the maximum branch divisions of the decision tree. At this stage, the division criterion was considered as "gdi". In terms of the parameter of this algorithm, the maximum number of splitting ("MaxNumSplits") was chosen as 20 according to node splitting rules [56].

*2.5.3. Ensemble aggregation 'bagging' and 'boosting'*

In the data analysis, the problem that classification studies often encounter is class irregularity and imbalance. Different methods have been developed to eliminate this problem, which occurs in disease diagnosis, face recognition, fluid leak diagnosis and many different areas [57]. Three different ways are followed to solve this problem. In the first way, by emphasizing the importance of positive samples, the new algorithm is obtained by correcting or changing the existing algorithms (i.e., algorithm level) [58]. In the second way, a pre-processing step is added. The aim is to minimize the effect of class distributions that suddenly change direction and decision (i.e., data level) [59]. In the third method, cost-sensitive methods are reinforced by combining algorithm and data level approaches [60].

In addition to the three approaches mentioned above, the ensembles technique is used in the class imbalance problem [61]. This algorithm, which is used in classification and regression in the field of artificial intelligence based on machine learning [62], is designed to increase the stability and accuracy of the existing algorithms [63]. Generally, this learning method performs discrimination by creating classifiers from the trained data and voting on the predictions obtained from each classifier [64]. In performing the ensembles method, it is important to choose classifiers that are compatible and consistent with the training set.

Bagging and boosting are the most common ensemble techniques that make big changes in a low performance and powerless classification algorithm [65], [66]. The biggest advantage of these techniques is to prepare the classifiers in advance depending on the desired variety while taking into account the training set.

The bootstrapping concept presented by Breiman for the bagging technique creates a new dataset for training classifiers by randomly drawing class samples. The basis of the method is to gain variety by re-sampling using data subsets. After obtaining this variety, the class of an unknown guest sample is determined by voting [57].

The boosting technique, recognized by Schapire in 1990, has proven to be authoritative to produce a powerful classifier model [67]. AdaBoost, one of the most effective and powerful members of the boosting family, is used as a representative approach in data mining [68]. By minimizing variance and deviation, it performs well via increasing the margin between the classes, similar to the working principle of the SVM. The working logic of AdaBoost can be summarized as follows: using the whole dataset, it makes a focus and effort to correctly guide the hard-to-separated samples by considering the samples misclassified in each iteration. It reduces the weights of the correctly classified samples and increases the weights of the wrong samples by playing with the sample weights in each iteration.

Different weights are assigned to each classifier during the testing phase. This process gives more credit and confidence to efficient classifiers. Thus, the choice of the class label for a guest instance is determined by the majority voting per classifier.

In this study, "AdaBoost.M1", which is a derivative of the boosting algorithm and the "bagging" technique were used to smooth out the class imbalance. The maximum number of splitting for the "bagging" and "AdaBoost.M1" algorithms was chosen as 3715 and 100, respectively.

## *3. Results*

In the MFCC algorithm, which is used as a feature extraction step in the data analysis phase, priority is given to calculating the Mel frequency cepstral coefficients for each classifier. In the M×N matrix obtained as the result of the MFCC application, the M and N parameters showed the Mel frequency coefficients and the number of windows, respectively. In the present study, this coefficient was evaluated between 2 and 39 for each classifier. A hamming window, similar to a raised cosine structure and without zero ends, was used in the study. Using trial-and-error theory, the length of this window and the window overlap were selected as (4×512) and (1024+512), respectively. As the result of the evaluation made between 2 and 39, the effective Mel coefficient value was determined as 23. After selecting the appropriate coefficient, seven statistical values for each coefficient index were considered as features: mean, standard deviation (SD), root means square (RMS), entropy, kurtosis (KUR), skewness (SKW) and variance (VAR). Thus, the size of the feature vector was calculated as M*7 (23*7=161). After the features were obtained, the accuracy rates of the classifiers on the validation set using different MFCC coefficients are presented in Fig. 2. The reason why the Mel coefficient was chosen as 23 was the maximum accuracy rate of the SVM algorithm, which provided successful performance, at this value.

In order to increase the performance of the study, the efficiency of classification algorithms was focused on by applying SFS feature selection. By performing the SFS method to each classifier and considering 23 effective Mel coefficients, the improved accuracy rates for SVM, decision tree, "AdaBoost.M1" and "bagging" are shown in Figs. 3, 4, 5 and 6, respectively. Owing to the feature selection used to optimize the number of features in the dataset, omit unnecessary data, reduce the training time and, most importantly, increase the accuracy rate [69], 80 outstanding and effective features specific to each classification were selected from among 161 features. A closer look at Fig. 2 shows that, among the classifiers, the SVM provided successful performance and the decision tree yielded less. In the SVM algorithm, which exhibited successful accuracy, the highest percentage of accuracy was shown as 80.31% in the Mel coefficient, which as 23 as seen in Fig. 2.

Considering the success of each classifier in Figs. 3, 4, 5 and 6, the effective number of features was chosen as 74, considering the behavior of the successful SVM classifier. As the result of this selection, according to Fig. 3, the accuracy success rate of SVM reached 83.18%. In other words, owing to the approximately 50% decrease in the number of total features (161), the classifier success percentage increased by 2.87%. This success was determined as 75.22%, 78.5% and 78.99% for the other three classifiers, as presented in Figs. 4, 5, and 6. On the other hand, when the graph of accuracy success according to 80 features in the SVM classifier was taken into visual analysis, the 83.01% success of this classifier in the number of 45 features cannot be overlooked. Also, it was observed that the ensemble aggregation derivatives 'bagging' and "AdaBoost-M1" showed similar behaviors.

The interpretation and final result of this analysis on the validation data are presented in Table 1. Based on the presented graphs, the successful accuracy rate in the selected effective 80 features were more pronounced for the first 74 features. Likewise, the accuracy and sensitivity values of the testing data for the proposed classifier algorithms after selecting effective features are presented in Table 2. As seen in these tables, in the coughing and sneezing sound analysis early diagnosis study, while the SVM algorithm appeared to be successful in the validation data, the "bagging" algorithm achieved high accuracy in the testing data.

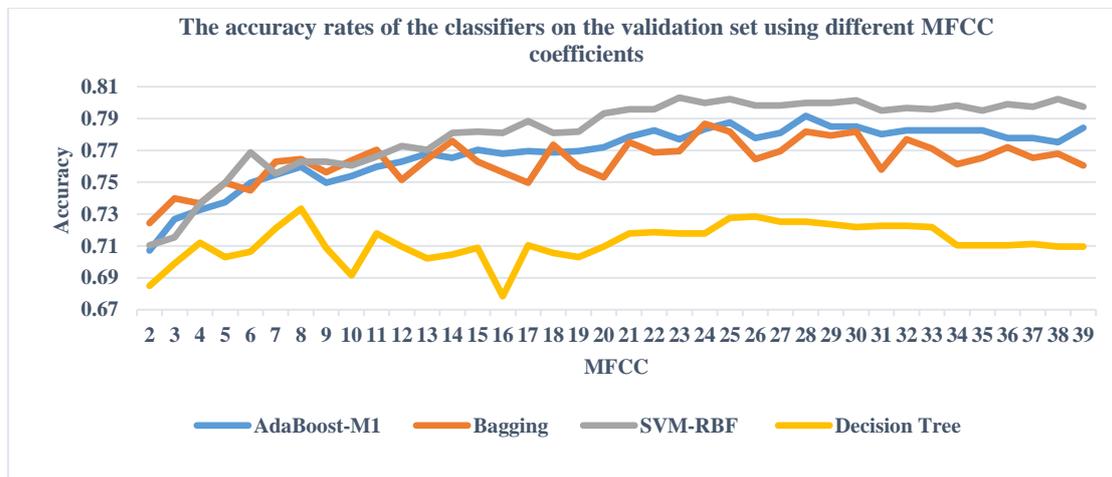

Fig. 2. Accuracy rates of the classifiers on the validation set using different MFCC coefficients.

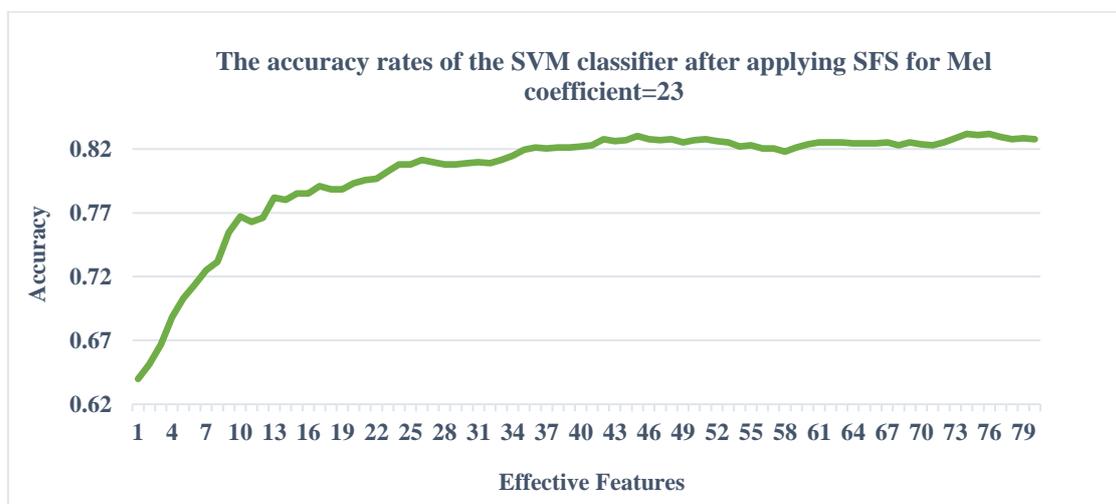

Fig. 3. Accuracy rates of the SVM classifier after applying SFS.

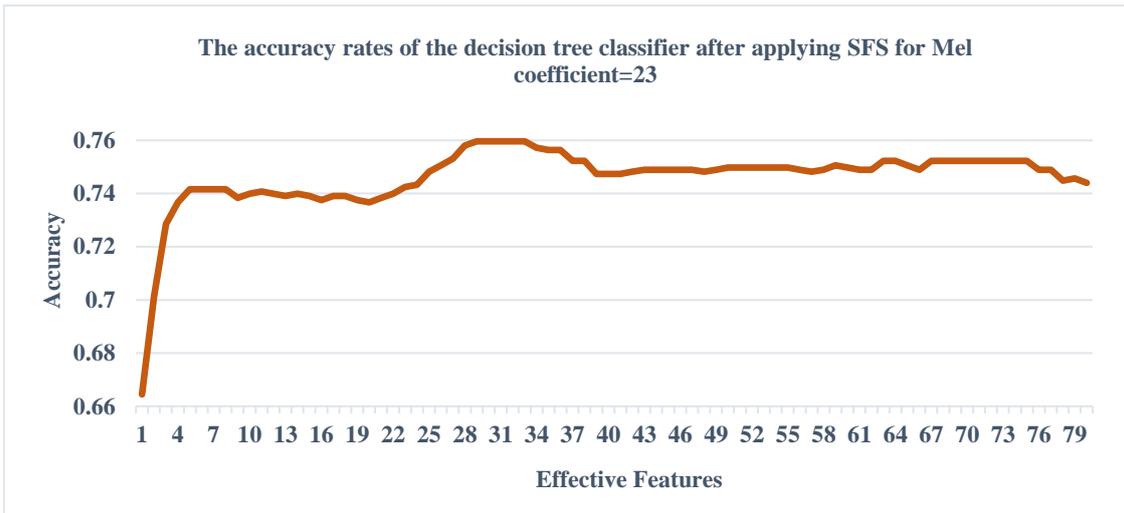

Fig. 4. Accuracy rates of the decision tree classifier after applying SFS.

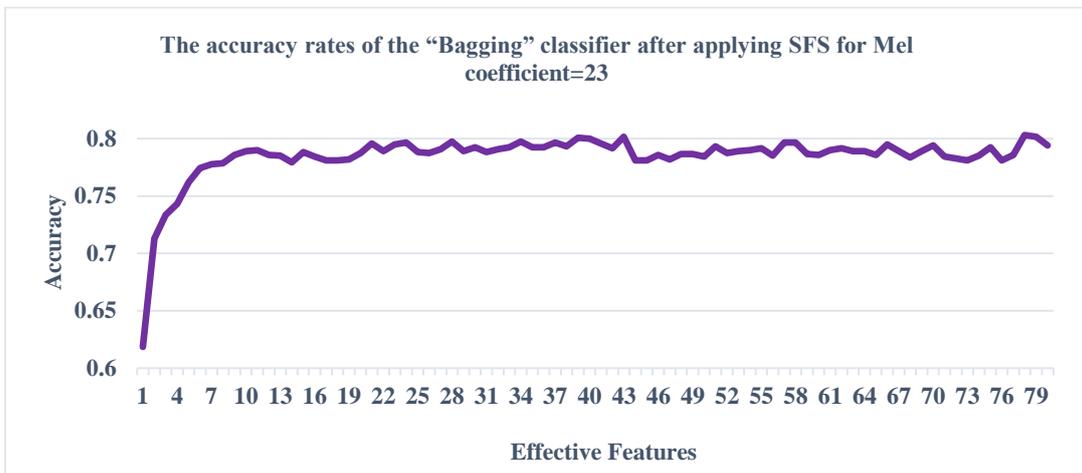

Fig. 5. Accuracy rates of the "Bagging" classifier after applying SFS.

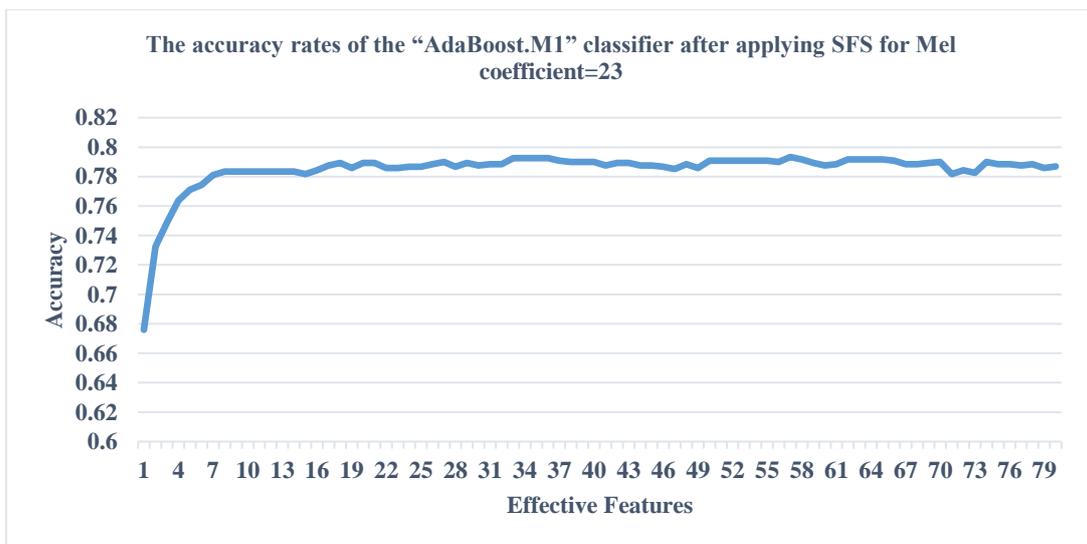

Fig. 6. Accuracy rates of the "AdaBoost.M1" classifier after applying SFS.

Table. 1. Classification accuracies and sensitivity values of validation data for 74 effective features and Mel coefficient=23.

|  | Classification parameters | Accuracy | Sensitivity-patient | Sensitivity-non patient |
|---|---|---|---|---|
| Decision tree | "MaxNumSplits"=100 | 0.7596 | 0.7752 | 0.7344 |
| AdaBoost.M1 | "MaxNumSplits"=20 | 0.8022 | 0.8603 | 0.7087 |
| Bagging | "MaxNumSplits"=3715 | 0.8014 | 0.8484 | 0.7259 |
| SVM | δ=1 | **0.8318** | 0.8776 | 0.7580 |

Table. 2. Classification accuracies and sensitivity values of testing data for 74 effective features and Mel coefficient=23.

|  | Classification parameters | Accuracy | Sensitivity-patient | Sensitivity-non patient |
|---|---|---|---|---|
| Decision tree | "MaxNumSplits"=100 | 0.7754 | 0.8180 | 0.7082 |
| AdaBoost.M1 | "MaxNumSplits"=20 | 0.7518 | 0.8367 | 0.6177 |
| Bagging | "MaxNumSplits"=3715 | **0.7784** | 0.8377 | 0.6848 |
| SVM | δ=1 | 0.7076 | 0.7240 | 0.6817 |

## *4. Discussion*

Detailed analysis of respiratory sounds opens up great possibilities in the medical world. Advance in technology and the reflection of this development on the medical world have led to the design of digital recording and advanced instruments [69], [70]. The recording and visualization of respiratory sounds, which have different characteristics, can be an early harbinger of some diseases that pose dangers to people's lives [71]. For this reason, a comprehensive classification study based on sneezing and coughing sounds has been conducted to analyze the diversity of sounds originating from the respiratory system and to think this analysis will be useful for the early diagnosis of diseases.

We found no comprehensive studies on the presented data set that responded to the "Pfizer digital medicine challenge" invitation. As mentioned, since there is no study on this data set, it was not considered appropriate to compare the results of the presented model with studies on different data sets. But in general, it seems useful to explain the results of some cough-based studies. It has been known for many years that coughing sound is a symptom of many diseases and has medical importance. Based on this information, a compilation study was made by Korbas et al. [72]. In a 2006 study, healthcare professionals analyzed cough sounds, showing that these analyzes would help them identify cough sound characteristics. However, it is seen that the analyses made in those years were insufficient in terms of diagnosis [73]. Wavelet analysis of volunteers with cough-based disease originating from the respiratory system was the main target of another study [74]. Using the discriminant analysis method, nearly 90% success was achieved by separating the cough sounds of healthy subjects from the cough sound characteristics of volunteers with asthma bronchial and chronic lung diseases. Irregular and abnormal pulmonary function detection was performed by classifying cough sound and air flow patterns. In this successful study, a new model for cough sound classification was developed. In this system, a bright light was shed in the medical world by diagnosing abnormal lung functions [75] via taking into account the acoustic properties of airflow and coughing sound [76]. The advantages and difficulties of cough sound-based studies in terms of disease diagnosis were examined in a comprehensive current review study [7]. Finally, we emphasized the importance of cough while taking into account the different studies reviewed over the last 25 years or so. From the early diagnosis of respiratory system diseases, we found that cough is a strong candidate as it exhibits variable patterns. Only the passage of years and advancement of technology seem to have the potential to turn this cough candidate into a more powerful and rapid diagnostic tool.

This study answered a challenge posed by the "OSFHOME" platform, proving that this dataset has sufficient potentials for disease early detection model design. A machine learning model was developed to detect sounds such as coughing and sneezing in this data set, which was a combination of ESC-50 and AudioSet audio files. Since 2016, different studies have been carried out on the ESC-50 data set. These studies with different objectives have generally focused on sound event recognition, sound classification with neural networks and environmental sound classification [23], [77], [78]. Statistical features were obtained by applying the MFCC feature extraction method, which is common in sound analysis studies [31], [79] in this three-stage data set. Using SFS method aimed to increase the performance of classification techniques by reducing the number of features to about half and selecting effective features. Coughing and sneezing sounds were successfully classified from other sounds in this mixed and crowded data set using three different classification algorithms. As the result of the selection of effective features, the SVM-RBF classifier was ahead of other classifiers with 83.18% success in the validation data. On the other hand, in the testing data, the bagging classifier, which is one of the ensemble aggregation algorithm family, achieved successful performance by distinguishing coughing and sneezing sounds from other sounds with 77.84% success.

This machine learning model based on coughing and sneezing sound analyzes seems to be useful for early diagnosis of diseases such as COVID-19 [80], [81], . The system can be used as an effective and distinctive tool in crowded environments by carrying the ability to install it on smart phones as an application. Since these models depend on automatic machine learning technology, they can minimize the risk of virus transmission in terms of human interaction in infectious diseases [82], [83].

## 5. Conclusion

A detailed analysis of respiratory system sounds such as coughing and sneezing was taken into account to accelerate the disease diagnosis. Owing to the design of the proposed automatic model, the diagnosis and treatment of respiratory system diseases that can lead to fatal consequences can accelerate. In this case, in crowded environments, a portable application or smart model design based on machine or deep learning is indispensable for this purpose. The proposed study was presented as a strong response to an invitation called the "Pfizer digital medicine challenge". In the three-stage dataset consisting of ESC-50 and AudioSet audio files, the features that best represent coughing and sneezing sounds were extracted using the MFCC method. After obtaining the statistically appropriate features based on this feature extraction method, the coughing and sneezing sound classification process was successfully performed for three different classifiers. Considering the results, SVM was the successful algorithm among the SVM, ensemble aggregation and decision tree classifiers. Although there was no detailed study on this dataset, the proposed model was useful in the automatic and early detection of coughing and sneezing sounds in terms of classification accuracy.

Future development of the study was planned as follows: tuning a larger dataset using different datasets, developing method comparison-based work by applying other feature extraction techniques and classification algorithms and testing up-to-date analytical tools applicable to large datasets.